\documentclass[12pt]{amsart}
\usepackage{amsmath}  
\newcommand{\cal}{\mathcal}

\newcommand{\lh}{{\cal L}({\cal H})}

\newcommand{\hi}{{\cal H}}
\newcommand{\ip}[2]{\left\langle\,#1\,|\,#2\,\right\rangle}
\newcommand{\ket}[1]{\mid#1\rangle}  
\newcommand{\kb}[2]{|#1\,\rangle\langle\,#2|} 
\newcommand{\fii}{\varphi}

\newcommand{\br}{\mathcal B(\mathbb R)}

\newtheorem{theorem}{Theorem}
\newtheorem{lemma}[theorem]{Lemma}
\newtheorem{remark}[theorem]{Remark}
\newtheorem{corollary}[theorem]{Corollary}
\newtheorem{proposition}[theorem]{Proposition}

\begin{document}

\title[Number-phase complementarity]{Are number and phase complementary observables?}

\author{P. Busch}
\address{Paul Busch, 
Department of Mathematics, University of Hull, Hull, England}
\email{p.busch@maths.hull.ac.uk}
\author{P. Lahti}
\address{Pekka Lahti,
Department of Physics, University of Turku, FIN-20014 Turku, Finland}
\email{pekka.lahti@utu.fi}
\author{J.-P. Pellonp\"a\"a}
\address{Juha-Pekka Pellonp\"a\"a,
Department of Physics, University of Turku, FIN-20014 Turku, Finland}
\email{juhpello@utu.fi}
\author{K. Ylinen}
\address{Kari Ylinen,
Department of Mathematics, University of Turku, FIN-20014 Turku, Finland}
\email{kari.ylinen@utu.fi}


\begin{abstract}
We study  various ways of characterising the quantum optical number  and phase as complementary observables.
\end{abstract}

\maketitle

\section{Introduction}
The aim of this paper is to clarify the sense in which number and phase
in quantum optics can be described as complementary observables.  
Here phase observables are characterised as phase shift covariant positive 
operator measures, 
with the number operator playing the part of the shift generator.

Let  $\hi$ be a complex separable Hilbert space, 
$(\ket n)_{n\geq 0}$ an orthonormal basis,
and $N=\sum_{n=0}^\infty n\kb nn $
the associated number observable with the  domain
$\mathcal D(N)=\{\fii\in\hi\,|\, \sum_{n\geq 0}n^2|\ip{\fii}{n}|^2<\infty\}$. 
Let $\lh$ denote the set of bounded operators on $\hi$, and let
$\mathcal{B}\left([0,2\pi)\right)$ denote the 
$\sigma$-algebra of the Borel subsets of the interval
$[0,2\pi)$. 
We define a phase observable
as a positive normalised operator measure
$\mathcal{B}\left([0,2\pi)\right)\ni X\mapsto E(X)\in\lh$
which is covariant under the  shifts  generated by the number observable:
\begin{equation}\label{covariance}
e^{ixN}E(X)e^{-ixN} = E(X+x)
 \end{equation}
for all $X\in \mathcal{B}\left([0,2\pi)\right)$ and  $x\in[0,2\pi)$,
where the addition $X+x$ is modulo $2\pi$.
The effects $E(X), X\in\mathcal{B}\left([0,2\pi)\right)$, are  then 
of the form \cite{Holevo, LP99, GEPJP}
\begin{equation}\label{ex}
E(X) = \sum_{n,m=0}^\infty c_{n,m}\, \frac{1}{2\pi}\int_Xe^{i(n-m)x}\mathrm dx\kb nm, \ X\in\mathcal{B}\left([0,2\pi)\right),
\end{equation}
where  
$(c_{n,m})_{n,m\geq 0}$ is the associated {\em phase matrix}, that is, a complex matrix generated by
a sequence of unit vectors $(\xi_n)_{n\geq 0}$ in $\hi$:
$c_{n,m}=\langle\xi_n|\xi_m\rangle$ for all $n,m\in\mathbb N$.

It is well known that among the phase observables 
there is no projection measure, that is, there is no self-adjoint operator whose spectral measure
would be phase shift covariant.

We proceed as follows. In Sec.\ 2 we present various distinct classes of phase observables, 
which will  provide examples illustrating the degree of commutativity of phase observables 
in Section 3 and the noncoexistence of number and phase in Section 4. Section 5 reviews the
different formalisations of complementarity, which are then applied to number-phase pairs in
Sections 6 and 7. 
It turns out that only such phase observables $E$, for which $\parallel E(X)\parallel =1$ for
$E(X)\ne O$, can be complementary to number. In Section 8 we show that the canonical
phase as well as the ground state phase space phase (the angle margin of the Husimi
$Q$-function) fulfill this necessary condition for the number-phase complementarity.
Finally, in Section 9 we address the question of the operational content of the
notions of complementarity studied here and in earlier work.


\section{Examples of phase observables}

The {\em canonical phase} is defined by the phase matrix whose entries are
$c_{n,m} =1$ for all $n,m\geq 0$:
\begin{equation}\label{can}
E_{\rm can}(X) = \sum_{n,m=0}^\infty \, \frac{1}{2\pi}\int_Xe^{i(n-m)x}\mathrm dx\kb nm, \ X\in\mathcal{B}\left([0,2\pi)\right).
\end{equation}
This 
is the unique semispectral measure associated with the polar decomposition of
the lowering operator, $a=\sum\sqrt{n+1}\kb n{n+1}$ $=V|a|=V\sqrt{N}$, $V=\int_0^{2\pi}e^{ix}\,dE_{\rm can}(x)$,
see, e.g. \cite[pp. 141-2]{Holevo82}, or \cite[Example 3.4]{LP99}.
With the choice of the identity matrix as the phase matrix one defines the {\em trivial phase},
\begin{equation}\label{triv}
E_{\rm triv}(X) =  \frac{1}{2\pi}\int_X\mathrm dx\,I = \frac{\ell(X)}{2\pi}\,I, \ X\in\mathcal{B}\left([0,2\pi)\right),
\end{equation}
where $\ell (X)$ denotes the Lebesgue measure of the Borel set $X$.
We define an {\em elementary phase} through the equation 
\begin{equation}\label{el}
E_{\rm el}(X) =  \frac{\ell(X)}{2\pi}\,I +  
w\,\frac{1}{2\pi}\int_Xe^{i(s-t)x}\mathrm dx\kb st
+ \overline{w}\,\frac{1}{2\pi}\int_Xe^{i(t-s)x}\mathrm dx\kb ts,
\end{equation}
where $s\ne t$ and $w$ is any complex number with $|w|\leq 1$.
Finally, the matrix elements
\begin{equation}\label{perustila}
c_{n,m}^{|0\rangle} = \frac{\Gamma(\frac{n+m}2+1)}{\sqrt{n!}\sqrt{m!}}, \ n,m\in\mathbb N,
\end{equation}
constitute the phase space phase observable generated by the ground state $|0\rangle$:
\begin{eqnarray*}
E_{|0\rangle}(X) &&= 
\sum_{n,m=0}^\infty c_{n,m}^{|0\rangle}\, \frac{1}{2\pi}\int_Xe^{i(n-m)x}\mathrm dx\kb nm\\
&&  = \frac{1}{\pi}\int_X\int_0^\infty|re^{i\theta}\rangle\langle re^{i\theta}|\,r{\rm d}r\,{\rm d}\theta,
\;\;\;X\in\mathcal{B}\left([0,2\pi)\right).
\end{eqnarray*}

We recall that phase observables $E_1$ and $E_2$ are {\em unitarily equivalent}
(as phase shift covariant observables)
if there is a unitary map $U=\sum_{n=0}^\infty e^{i\vartheta_n}\kb nn$ such that
$E_2(X) = UE_1(X)U^{-1}$ for all $X\in\mathcal{B}\left([0,2\pi)\right)$.
A phase observable $E$ is called {\em strong} if its $k$-th cyclic moment operator
$V_E^{(k)} :=\int_0^{2\pi}e^{ikx}\mathrm dE(X)$  is the $k$-th power of its
first cyclic moment operator $V_E^{(1)}$, that is, if for each $k\geq 0$,
$V_E^{(k)}=(V_E^{(1)})^k$. 
If $E$ is strong, then $\mathbb N\ni k\mapsto V_E^{(k)}\in\lh$ constitutes
a (nonunitary) representation of the additive semigroup of the nonnegative
integers, and one may ask whether the
number observable $n\mapsto \kb nn$ behaves covariantly under the one-sided shifts 
generated by the phase.
Observe that $V_E^{(k)}\ket{n+k}=c_{n,n+k}\ket{n}$ for all $n,k\in\mathbb N$
which shows that $N$ is $E$-covariant whenever $|c_{n,m}|=1$ for all 
$n,m\in\mathbb N$.

Combining results from \cite{LP99,LP00} we have the following theorem:

\begin{theorem}\label{uni}
For any phase observable $E$, with the phase matrix $(c_{n,m})$, the following conditions
are equivalent:
\begin{itemize}
\item[(a)]
$E$ is unitarily equivalent with $E_{\rm can}$;
\item[(b)]
$|c_{n,m}|=1$ for all $n,m$;
\item[(c)]
$E$ generates the number shifts.
\end{itemize}
\end{theorem}

\section{The degree of commutativity of a phase observable}

Let ${\rm com}\,(E)$ denote the set of vectors $\fii\in\hi$ for which
$$
E(X)E(Y)\fii = E(Y)E(X) \fii \ {\rm for\ all } \ X,Y\in\mathcal B\left([0,2\pi)\right).
$$
We say that
$E$ is {\em commutative} if  ${\rm com}\,(E)=\hi$, and 
{\em totally noncommutative} if ${\rm com}\,(E) =\{0\}$. 

\begin{proposition}
A phase observable $E$ is commutative if and only if it is the trivial phase $E_{{\rm triv}}$.
\end{proposition}

\begin{proof}
The trivial phase $E_{{\rm triv}}$ is commutative. 
Let $E$ be a phase observable with the matrix $(c_{n,m})_{n,m\geq 0}$.
For any $n\in\mathbb N$ and $Y\in \mathcal B\left([0,2\pi)\right)$, the map
$$
\mathcal B\left([0,2\pi)\right)\ni X\mapsto \mu_{n,Y}(X) := \ip{n}{(E(X)E(Y)-E(Y)E(X))|n}\in\mathbb C
$$
is  a complex measure.  For any $k\in\mathbb N$,
\begin{equation*}
\int_0^{2\pi} e^{ikx}\mathrm d\mu_{n,Y}(x) =
\left\{ 
\begin{array}{ccc}
|c_{n,n+k}|^2 \frac 1{2\pi}\int_Y e^{ikx}\,\mathrm dx & & \ {\rm when } \ n<k,\\
&&\\
\left(|c_{n,n+k}|^2-|c_{n-k,n}|^2\right) \frac 1{2\pi}\int_Y e^{ikx}\,\mathrm dx
& & \ {\rm when } \ n\geq k.
\end{array}
\right.
\end{equation*}
If  $E$ is commutative, then
$\int_0^{2\pi} e^{ikx}\mathrm d\mu_{n,Y}(x)=0$ for all $Y\in \mathcal B\left([0,2\pi)\right)$, 
so that  $c_{n,n+k}=0$, for $n<k$, and $|c_{n,n+k}|=|c_{n-k,n}|$, for $n\geq k$.
Let $n\geq k$, and let $l\geq 0$ be the smallest integer for which $n'\equiv n-lk<k$.
Then $|c_{n,n+k}|=|c_{n-k,n}|= \cdots =|c_{n',n'+k}|=0$. But this means that $c_{n,m}=0$ for all
$n\ne m$, that is, $E=E_{{\rm triv}}$.
\end{proof}

\begin{lemma}\label{come1}
Let $E$ be a phase observable with the matrix $(c_{n,m})_{n,m\geq 0}$.  Then
\begin{equation}\label{come}
\left\{ \fii\in\hi\,|\, \ip{n}{\fii}=0\ {\rm if}\ c_{n,m}\ne 0\ {\rm for\ some}\ m\ne n \right\}
\subseteq {\rm com}\,(E) .
\end{equation}
\end{lemma}

\begin{proof}
For the phase observable $E$ with the matrix $(c_{n,m})_{n,m\geq 0}$, let
$$
A = \{n\in\mathbb N\,|\, c_{n,m}=0 \ {\rm for\  all}\ m\ne n\},
$$
and define
$$
P_A = \sum_{s\in A} \kb ss.
$$
Then
$$
E(X)P_A=\frac{\ell(X)}{2\pi}P_A
$$
and therefore
$$
E(X)E(Y)P_A=\frac{\ell(X)\ell(Y)}{4\pi^2}P_A=E(Y)E(X)P_A
$$ 
for all $X,\,Y\in\mathcal B\left([0,2\pi)\right)$, so that
\begin{eqnarray*}
P_A  (\hi)&=&
 \{ \fii\in\hi\,|\, 
\ip{n}{\fii}=0\ {\rm if}\ c_{n,m}\ne 0\ {\rm for\ some}\ m\ne n \}\\
&\subseteq& 
{\rm com}\,(E).
\end{eqnarray*}
\end{proof}
There are phase observables for which the set inclusion (\ref{come}) is proper.
For instance, the phase observable $E_w$, with 
$c_{0,1}=\overline{c_{1,0}}=c_{1,2}=\overline{c_{2,1}}=w$, 
$w\in\mathbb C, 0<|w|\leq 1/\sqrt 2$,
and $c_{n,m}=0$ for all other $n,m\geq 0, n\ne m$, is such.


\begin{proposition}\label{strong}
Let $E$ be a phase observable with the matrix $(c_{n,m})_{n,m\geq 0}$. If $E$ is strong, then
\begin{equation}\label{comeq}
{\rm com}\,(E)  = \left\{ \fii\in\hi\,|\, 
\ip{n}{\fii}=0\ {\rm if}\ c_{n,m}\ne 0\ {\rm for\ some}\ m\ne n \right\}.
\end{equation}
\end{proposition}

\begin{proof}
Consider a vector  $\psi=\sum_{s=0}^\infty d_s\ket s\in\mathcal H$.
In view of Lemma \ref{come1},
it remains to be shown that $\psi\in{\rm com}\,(E)$
implies $\psi\in P_A  (\hi)$.
For any $n\in\mathbb N$ and $X,Y\in \mathcal B\left([0,2\pi)\right)$ we define
$$
F_{n,\psi}(X,Y) := \ip{n}{(E(X)E(Y)-E(Y)E(X))\psi}.
$$
For a fixed $Y$, the partial map $X\mapsto F_{n,\psi}(X,Y)$ is a complex measure. 
For any $k\in\mathbb N$,
\begin{eqnarray*}
&& \int_0^{2\pi} e^{ikx}\mathrm dF_{n,\psi}(x,Y)\\ &&=
\sum_{l=0}^\infty \left(c_{n,n+k}\ip{n+k}{E(Y)|l}d_l 
-     c_{l,l+k}\ip{n}{E(Y)|l}d_{l+k}  \right)\\
&&=: \tilde F_{n,k,\psi}(Y).
\end{eqnarray*}
Again, the map $Y\mapsto  \tilde F_{n,k,\psi}(Y)$ is a complex measure, and we may
carry out the integration
$$
\int_0^{2\pi} e^{iry}\mathrm d\tilde F_{n,k,\psi}(y)
=
\sum_{l=0}^\infty \left(c_{n,n+k}c_{n+k,l}d_l\delta_{0,n+k-l+r} 
-  c_{n,l} c_{l,l+k}d_{l+k}\delta_{0,n-l+r}  \right)
$$
for all $r\in\mathbb Z$. If $\psi\in {\rm com}\,(E)$, then
the value of the above integral is zero for all $n,k\in\mathbb N, r\in\mathbb Z$.
This implies 
\begin{eqnarray*}
c_{n,n+k}c_{n+k,n+k+r}d_{n+k+r} 
&=&  c_{n,n+r} c_{n+r,n+r+k}d_{n+r+k},\ {\rm when }\ r\geq -n,\\
c_{n,n+k}c_{n+k,n+k+r}d_{n+k+r}&=&0,\ {\rm when }\ -n>r\geq -n-k.
\end{eqnarray*}
On writing $m=n+k+r$ these conditions are equivalent to:
\begin{eqnarray*}
c_{n,n+k}c_{n+k,m}d_{m} 
&=&  c_{n,m-k} c_{m-k,m}d_{m},\ {\rm when }\ m\geq k,\\
c_{n,n+k}c_{n+k,m}d_{m}&=&0\ {\rm when },\ k>m\geq 0.
\end{eqnarray*}
It remains to be shown that if $d_m\ne 0$, then $c_{n,m}=0$ for all $n\ne m$.
Assume, therefore, that $d_m\ne 0$ for some $m\in\mathbb N$. Then
\begin{eqnarray}
c_{n,n+k}c_{n+k,m} 
&=&  c_{n,m-k} c_{m-k,m},\ {\rm when }\ m\geq k,\\
c_{n,n+k}c_{n+k,m}&=&0,\  {\rm when }\ k>m\geq 0.\label{9}
\end{eqnarray}
Putting $n=m$, we get $c_{m,m+k}=0$ for all $k>m$.
If $m=0$, then $c_{0,k}=0$ for all $k>0$.
Assume next that $m\ne 0$. Since $E$ is strong, we get
$$
c_{m,m+k}=c_{m,m+1}c_{m+1,m+2}\cdots c_{m+k-1,m+k}
$$ 
for all $k>0$ and
$$
c_{m,m-k}=\overline{c_{m-k,m}}= \overline{c_{m-k,m-k+1}}\overline{c_{m-k+1,m-k+2}}
\cdots\overline{c_{m-1,m}}
$$
for all $k>0$, $k\le m$.
Further, for $k=1$ and $n=m$, we get from the first of the above equalities that
$|c_{m-1,m}|=|c_{m,m+1}|$. Thus it suffices to show that $c_{m-1,m}=0$. For that,
assume the contrary: $c_{m-1,m}\ne 0$.  Fix $l=1,2,...$, and take $n=m+l,$ $k=1$. Then
$$
c_{m+l,m+l+1}c_{m+l+1,m}  = c_{m+l,m-1} c_{m-1,m}.
$$
Since $E$ is strong, this equation is equivalent to the following:
$$
|c_{m+l,m+l+1}|^2c_{m+l,m}  = |c_{m-1,m}|^2 c_{m+l,m}.
$$ 
Therefore, if $c_{m,m+l}\ne 0$, then also $c_{m+l,m+l+1}\ne0$, and
$$
c_{m,m+l+1}=c_{m,m+l}c_{m+l,m+l+1}\ne 0.
$$
We have thus shown that $c_{m,m+l}\ne 0$ for all $l=1,2,...$.
But this is impossible since, by choosing $n=m$ and $k=m+1$ in (\ref{9}), 
we have $c_{m,m+m+1}=0$.
Therefore, $c_{m-1,m}=0$. 
\end{proof}

We note that the equality (\ref{comeq}) is not restricted to strong phase observables only.
Indeed, an elementary phase with $c_{0,2}=c_{2,0}=1$ is not strong but it has the property (\ref{comeq}).
The canonical phase and all phase observables unitarily equivalent to it 
are strong, and all their matrix elements are of modulus one. Therefore, we have:

\begin{corollary}
The canonical phase, as well as any  phase observable unitarily equivalent to it,  is totally noncommutative.
\end{corollary}






\section{The noncoexistence of number and phase}

The notion of coexistence of observables has been introduced  to describe the
possibility of measuring the observables together (see, e.g., \cite{Ludwig, Kraus, QTM, LPY98}). 
If two observable are noncoexistent, they cannot be measured together. 
Since the number observable $N$
is given by a projection measure $n\mapsto \kb nn\equiv P_n$, 
the coexistence of the number $N$ and  a phase $E$ implies
their commutativity:
\begin{equation}\label{com1}
P_nE(X) =E(X)P_n\ \ \ {\rm for\  all }\ n\in\mathbb N, X\in\mathcal B\left([0,2\pi)\right).
\end{equation}
If we assume that the number $N$ and a phase  $E$ commute, then it immediately follows that  
\begin{equation}\label{com2}
E(X) = \sum_{n=0}^\infty P_nE(X)P_n=\sum_{n=0}^\infty\ip{n}{E(X)|n}P_n = \frac{\ell(X)}{2\pi}\,I
\end{equation}
for all $X$, which shows that $E$ is the trivial phase.
Hence we have:

\begin{corollary} 
Any nontrivial phase and number are noncoexistent observables.
\end{corollary}


Let ${\rm com}(N,E)$ denote the set of vectors 
$\fii\in\hi$ for which 
\begin{equation}\label{dcom}
P_nE(X)\fii =E(X)P_n\fii\ \ \ {\rm for\  all }\ n\in\mathbb N,X\in\mathcal B\left([0,2\pi)\right).
\end{equation}
Assume that $\fii\in{\rm com}(N,E)$. Then $P_nE(X)\fii = \frac{\ell(X)}{2\pi}P_n\fii$ 
for all $n$ and $X$ and thus $E(X)\fii = \frac{\ell(X)}{2\pi}\fii$ for all $X$.
Hence, if ${\rm com}(N,E)=\hi$, then $E$ is the trivial phase.
Moreover, if $\fii\in{\rm com}(N,E)$, $\fii\ne 0$, then $\ip{k}{\fii}\ne 0$ for some $k$,
in which case  $c_{n,k}=0$ for all $n\ne k$.  Therefore,
\begin{equation}\label{comne}
{\rm com}(N,E) =\{ \fii\in\hi\,|\, 
\ip{n}{\fii}=0\ {\rm if}\ c_{n,m}\ne 0\ {\rm for\ some}\ m\ne n \}.
\end{equation}
When combined with Lemma \ref{come1}, this shows that  
\begin{equation}\label{comneg}
{\rm com}(N,E)\subseteq {\rm com}\,(E).
 \end{equation}
There are examples of phase observables 
for which this inclusion is a proper one.  However,  
by Proposition \ref{strong}, if $E$ is strong, then
${\rm com}(N,E)={\rm com}\,(E)$.
In particular, we have:
\begin{proposition}
For any phase observable $E$,  if $c_{n,m}\ne 0$ for all $n,m\geq 0$, then
${\rm com}(N,E)=\{0\}$. 
\end{proposition}
\noindent
Thus, for instance, the phase observables $E_{\rm can}$ and $E_{|0\rangle}$ commute with
$N$ in no state.

We recall that for any $\fii\in{\rm com}(N,E)$, the map 
\begin{equation}\label{jp}
(n,X)\mapsto\ip{\fii}{P_nE(X)\fii}= |\ip{\fii}{n}|^2\frac{\ell(X)}{2\pi}
\end{equation}
is a probability bimeasure and thus
extends to a joint probability measure of number $N$ and phase $E$ in the vector
state $\fii$,  see,  e.g. \cite{BCR,KY}.


\section{Forms of complementarity}

The operational idea of complementarity of two observables 
in the sense of the mutual exclusion of any two experimental procedures 
permitting  the unambiguous definition of these quantities \cite{Bohr}
leads in the frame of the quantum theory of measurement \cite{QTM}
to the following condition on number and phase:
the number $N$ and a phase $E$ are {\em complementary}  if for any 
finite set $\{n_1,\cdots,n_k\}\subset\mathbb N$ 
and for any $X\in\mathcal B\left([0,2\pi)\right)$, for which  $O\ne E(X)\ne I$, 
\begin{equation}\label{c1}
(\sum_{i=1}^kP_{n_i})\land E(X)=O.
\end{equation}  

The probabilistic idea of complementarity of two observables
in the sense of mutual exclusion of the certain (probability one) predictions
of the values of the two observables leads, in the case of number and phase
to the following definition:
$N$ and $E$ are {\em probabilistically complementary} if
\begin{eqnarray}
 \sum_{i=1}^k |\ip{\fii}{n_i}|^2&=&1\ {\rm  implies\ that }\  0<\ip{\fii}{E(X)\fii}<1,\\
 \ip{\fii}{E(X)\fii}&=&1\ {\rm implies\ that }\   
0<\sum_{i=1}^k |\ip{\fii}{n_i}|^2<1,
\end{eqnarray}
for all vector states $\fii$, 
all nonempty sets $\{n_1,\cdots,n_k\}\subset\mathbb N$, 
and for any $X\in\mathcal B\left([0,2\pi)\right)$, such that $O\ne E(X)\ne I$.

There is yet another intuitive notion of complementarity, to which we refer
as value complementarity.  The idea is that if one of the  observables assumes a sharp
value, then the other should be uniformly distributed. There are technical problems
in formalising this idea because continuous quantities do not have eigenvalues
and for unbounded value sets a uniform distribution cannot be easily defined.
Here is our proposed definition :
$N$ and $E$ are {\em value complementary} if the following conditions are satisfied:
\begin{itemize}
\item[(i)]
 for any number eigenstate $\ket n$, the phase distribution $X\mapsto \ip{n}{E(X)|n}$ is uniform;
\item[(ii)]
for any sequence of vector states $(\fii_r)$, if the phase distributions  $X\mapsto \ip{\fii_r}{E(X)\fii_r}$
approach a delta distribution centred at some $x_0\in[0,2\pi)$, then the number distributions
$n\mapsto |\ip{\fii_r}{n}|^2$ get increasingly uniform, i.e., 
$|\ip{\fii_r}{n}|^2\to 0$ as $r\to\infty$. 
\end{itemize}

The probability distribution of any phase observable $E$ is uniform in every number state $\ket n$
as $\ip{n}{E(X)|n} = \frac{\ell(X)}{2\pi}$. Thus (i) is always fulfilled and
the value complementarity of $N$ and $E$ depends only on (ii).

\section{An example: the noncomplementarity of number and elementary phase}\label{esim.}

We consider the elementary phase $E_{\rm el}$ of Equation (\ref{el}). The spectrum of any 
$E_{\rm el}(X)\ne O,I$ consists of three eigenvalues
\begin{equation}\label{em0p}
0\leq e_-(X)\leq e_0(X)= \frac{\ell(X)}{2\pi}\leq e_+(X)\leq 1
\end{equation}
with
\begin{equation}\label{epm}
e_\pm(X) =\frac{\ell(X)}{2\pi}\pm |z|\,\left|\frac{1}{2\pi}\int_Xe^{i(s-t)x}\mathrm dx\right|.
\end{equation}
We note that $e_+(X) =1$ only when $E_{\rm el}(X)=I$, and $e_-(X) =0$ only when $E_{\rm
el}(X)=O$. Hence we have
\begin{equation}\label{norm}
\parallel{E_{\rm el}(X)}\parallel= e_+(X)<1 \ {\rm whenever\ } E_{\rm el}(X)\ne I, 
\end{equation}
and therefore, for any unit vector $\fii$,
\begin{equation}\label{prob}
\ip{\fii}{E_{\rm el}(X)\fii}<1\ {\rm whenever\ }E_{\rm el}(X)\ne I. 
\end{equation}

We recall that any operator $A$, with $O\leq A\leq I$, can be written in the form
$A=\lor(P\land A\,|\, P\ {\rm one\ dimensional\ projection})$ \cite{BuschGudder}, 
where $P\land A = \lambda P$, with 
$\lambda = \parallel A^{-1/2}\fii\parallel^{-2}$,
for any unit vector $\fii \in P(\hi)\cap {\rm ran}(A^{1/2})$ if the intersection is not the null space,
and $\lambda =0$ otherwise.
For any $E_{\rm el}(X)\ne O,I$ we now get
\begin{eqnarray*}
P_k\land E_{\rm el}(X) &=& \lambda(X)P_k,\ k=s,t ,\\  
\lambda(X) &=& 2\frac{e_-(X)e_+(X)}{e_-(X)+e_+(X)}\\
P_n\land E_{\rm el}(X)  &=& \frac{\ell(X)}{2\pi}P_n, \ n\ne s,t.
\end{eqnarray*}
These relations show that number and elementary phase are not complementary. 

For the probabilistic complementarity of $N$ and $E_{\rm el}$ we need to check only
the first implication in the definition (since the other holds trivially due to 
Eq. (\ref{prob})).
Assume that  $\sum_{i=1}^k |\ip{\fii}{n_i}|^2=1$. 
If $\{s,t\}\not\subset \{n_1,\cdots,n_k\}$, then 
$$
\ip{\fii}{E_{\rm el}(X)\fii}= \frac{\ell(X)}{2\pi}, 
$$
which is less than 1 whenever $E_{\rm el}(X)\ne I$. If $s,t\in \{n_1,\cdots,n_k\}$, then
\begin{eqnarray*}
\ip{\fii}{E_{\rm el}(X)\fii} &=& \frac{\ell(X)}{2\pi}+ 2\,{\rm Re}\,\left(z\ip{\fii}{s}\ip{t}{\fii}
\frac{1}{2\pi}\int_Xe^{i(s-t)x}\,\mathrm dx  \right)\\
&\leq& 
 \frac{\ell(X)}{2\pi} +  |z|\frac{\ell(X')}{2\pi},
\end{eqnarray*}
which is less than 1 whenever $|z|<1$ and  $E_{\rm el}(X)\ne I$. Therefore, $N$ and $E_{\rm el}$
are probabilistically complementary. In view of (\ref{prob}), they are also
value complementary.

\section{Number and canonical phase}

\begin{proposition}\label{norm}
The canonical phase satisfies
$\ip{\fii}{E_{\rm can}(X)\fii} < 1$ for any $X$ such that $E_{\rm can}(X)\ne I$ and for any unit vector $\fii\in\hi$.
\end{proposition}
\begin{proof}
The minimal Neumark dilation $\tilde{E}_{\rm can}$ of $E_{\rm can}$ in $L^2\left([0,2\pi)\right)$
is the canonical spectral measure $X\mapsto \tilde{E}_{\rm can}(X)$, 
with $\tilde{E}_{\rm can}(X)$  acting as multiplication by the characteristic function $\chi_X$.
The Hilbert space $\hi$ is identified with the Hardy space $H^2$ in $L^2\left([0,2\pi)\right)$.
If $\ip{\fii}{E_{\rm can}(X)\fii}=1$ for some unit vector $\fii\in H^2$ and for some $X$ 
for which $E_{\rm can}(X)\ne I$,
then $\fii$ vanishes on the complement set $X'$ which has positive measure. It follows 
from \cite[Theorem 13.13]{Young} that $\fii$ is zero, which is a contradiction.
\end{proof}

\begin{corollary}
Number and canonical phase are probabilistically complementary.
\end{corollary}

We consider next the value complementarity of $N$ and an arbitrary phase $E$.

\begin{proposition}
Let $(\psi_m)_{m\in\mathbb N}$ be a sequence of unit vectors for which the probability measures
$X\mapsto\ip{\psi_m}{E(X)\psi_m}$  tend (with $m\to\infty$) to a Dirac measure
$\delta_\theta$, $\theta\in [0,2\pi)$.
Then the number probabilities $|\ip{\psi_m}{n}|^2$ tend to zero for all $n$.
\end{proposition}

\begin{proof}
For the phase observable  $E$ with the phase matrix $(c_{n,m})_{n,m\in\mathbb
N}$, put
$p^E_\psi(X):=\langle\psi|E(X)\psi\rangle$, for any unit vector $\psi$.
Let $(\psi_m)_{m\in\mathbb N}\subset\mathcal H$ be a sequence of unit vectors
such that 
$$
\lim_{m\to\infty}p^E_{\psi_m}([0,x))=\delta_\theta([0,x))=
\left\{\begin{array}{ll}0
& \textrm{when $0<x<\theta$}\\
1
& \textrm{when $\theta<x\le 2\pi$}
\end{array}\right.\\
$$ 
where 
$\delta_\theta$ is the Dirac measure concentrated on the point $\theta\in [0,2\pi)$.
This implies that
$\lim_{m\to\infty}\int_0^{2\pi}e^{ikx}\mathrm dp^E_{\psi_m}(x)
=e^{ik\theta}$ for all $k\in\mathbb N$ (see e.g.\ \cite[Theorem 26.3]{Bi}). Then
$$
\left|\int_0^{2\pi}e^{ikx}\mathrm dp^E_{\psi_m}(x)\right|^2
=\left|\sum_{n=0}^\infty 
c_{n,n+k}\overline{\langle n|\psi_m\rangle}\langle n+k|\psi_m\rangle\right|^2
\le \sum_{l=k}^\infty\left|\langle l|\psi_m\rangle\right|^2 \le 1.
$$
The left hand side converges to 1 for all $k\in\mathbb N$, and so
$$
\sum_{l=k}^\infty\left|\langle l|\psi_m\rangle\right|^2 \to 1\ {\rm as\ }m\to\infty
$$
for all $k\in\mathbb N$ if and only if
\begin{equation}\label{raja}
\sum_{n=0}^p\left|\langle n|\psi_m\rangle\right|^2\to 0\ {\rm as\ }m\to\infty
\end{equation}
for all $p\in\mathbb N$.
\end{proof}
Equation (\ref{raja}) implies that $\lim_{m\to\infty}\langle\psi_m|N\psi_m\rangle\to\infty$.
This situation, where the number gets large and the phase arbitrarily
well defined, corresponds to the classical limit for a single
mode photon field.

\begin{corollary}
Number $N$ and any phase $E$ are value complementary.
\end{corollary}

\section{On the norm of the phase effects $E(X)$}

In Section \ref{esim.} we saw that the norm of the effects $E_{\rm el}(X)$ is strictly less than
one whenever $E_{\rm el}(X)\ne I$. On the other hand,
a phase $E$ can be complementary to  number only if
$\parallel E(X)\parallel =1$ for  all $E(X)\ne O$. Indeed, assume that $\parallel E(X)\parallel <1$
for a nonzero effect $E(X)$. Then the range of $E(X')$ is $\hi$ and therefore $P\land E(X')\ne O$
for any one dimensional projection $P$. Hence $N$ and $E$ are noncomplementary.
In this section we show that the norms of any  nonzero effects of $E_{\rm can}$
and $E_{|0\rangle}$ are one. 
To prove this claim   we need to develop some auxiliary results.

To start with we recall (e.g. from \cite[p. 138]{Rudin}) that a point $x\in\mathbb R$ is a {\em Lebesgue point}
of a Lebesgue integrable function $f:\mathbb R\to\mathbb R$, if
$$
\lim_{r\to0_+}\frac 1{2r}\int_{[x-r,x+r]}|f(y)-f(x)|\,dy =0.
$$
We only need this notion in the case where $f$ is the characteristic function $\chi_X$ of a 
Borel 
$X$. Clearly, $x\in X$ is a Lebesgue point of $\chi_X$ if and only if
$$
\lim_{r\to0_+}\frac 1{2r}\,{\ell}(X\cap[x-r,x+r])=1.
$$

\begin{lemma}\label{lemmaa}
Let $X\in\br$
and $x\in X$ a Lebesgue point of $\chi_X$.
Then
$$
\lim_{r\to0_+}\frac 1{r}\,{\ell}(X\cap[x-r,x])=1,
$$
and
$$
\lim_{r\to0_+}\frac 1{r}\,{\ell}(X\cap[x,x+r])=1.
$$
\end{lemma}

\begin{proof}
We prove the first equality; the proof of the second is similar.
Let $\epsilon >0$. There is a $\delta >0$ such that ${\ell}(X\cap[x-r,x+r])\geq 2r-\epsilon r$,
whenever $r<\delta$. Since ${\ell}(X\cap(x,x+r])\leq r$, it follows that
$\ell(X\cap [x-r,x])={\ell}(X\cap[x-r,x+r])-{\ell}(X\cap(x,x+r])\geq 2r-\epsilon r - r
= r(1-\epsilon)$ if $0<r<\delta$.
\end{proof}

\begin{lemma}\label{lemmab}
Let $f:[0,\infty)\to[0,\infty)$ be a (not necessarily strictly) decreasing function, and 
let $X\in\br$. 
Suppose that there are
numbers $q\in[0,1]$ and $\delta>0$ such that
${\ell}([0,r]\cap X) \geq rq$, whenever $0<r\leq \delta$. Then
$$
\int_{X\cap[0,\delta]}f(x)\,{\rm d}x \geq q\int_{[0,\delta]}f(x)\,{\rm d}x.
$$
\end{lemma}

\begin{proof}
By changing the values of $f$ on a countable set if necessary, we may assume that $f$ is
left continuous. Denote $a=f(0)-f(\delta)$. If $n\in\mathbb N$, for $k=0,1,\cdots,n$ write
$$
x_k = \sup\left\{x\in[0,\delta]\,\Big|\, f(x)\geq f(0)-\frac kna\right\},
$$
so that $0\leq x_0\leq x_1\leq\cdots\leq x_n=\delta$.
If $x_k\leq t\leq x_{k+1}$, then
$$
f(0)-\frac kna\geq f(t)\geq f(0)-\frac{k+1}na,
$$
and it is easily seen that
$$
\delta f(\delta)+\frac an\sum_{k=1}^nx_k\geq\int_{[0,\delta]}f(x)\,{\rm d}x\geq
\delta f(\delta)+\frac an\sum_{k=0}^{n-1}x_k.
$$
Since the difference of the left and right extremes is $\frac an\delta$, we see that for any
$\epsilon>0$, $n$ can be chosen such that $\frac an\delta<\epsilon$ and thus
$$
\delta f(\delta)+\frac an\sum_{k=0}^{n-1}x_k +\epsilon \geq\int_{[0,\delta]}f(x)\,{\rm d}x\geq
\delta f(\delta)+\frac an\sum_{k=0}^{n-1}x_k.
$$
Let $\ell_2$ be the two-dimensional Lebesgue measure and denote
$$
Z_k = [0,x_k]\times\left[f(0)-\frac{k+1}na,f(0)-\frac kna\right]
$$
for $k=0,\cdots,n-1$, and $Z_n=[0,\delta]\times[0,f(\delta)]$.
Then by the assumption we get
\begin{eqnarray*}
\int_{X\cap[0,\delta]}f(x)\,{\rm d}x
&&\geq\sum_{k=0}^n\ell_2(Z_k\cap\{(x,y)\,|\, x\in X\})\\
&&=f(\delta)\ell([0,\delta]\cap X)+\frac an\sum_{k=0}^{n-1}\ell([0,x_k]\cap X)\\
&&\geq q\left[\delta f(\delta)+\frac an\sum_{k=0}^{n-1}x_k\right]
\geq q\int_{[0,\delta]}f(x)\,{\rm d}x-q\epsilon.
\end{eqnarray*}
Letting $\epsilon\to 0$ we get the claim.
\end{proof}

\begin{theorem}\label{lausea}
Let $(f_n)_{n\in\mathbb N}$ be a sequence of functions $f_n:\mathbb R\to[0,\infty)$ such that
\begin{itemize}
\item[(i)]
$f_n(x)\leq f_n(y)$, if $x\leq y\leq 0$;
\item[(ii)]
$f_n(x)\geq f_n(y)$, if $0\leq x\leq y$;
\item[(iii)]
$\int_{\mathbb R} f_n(x)\,{\rm d}x = 1$;
\item[(iv)]
$\lim_{n\to\infty}\int_{[-\delta,\delta]}f_n(x)\,{\rm d}x =1$ for any $\delta>0$.
\end{itemize}
(a) If $X\in\br$ 
is such  that $0\in X$ and $0$ is a 
Lebesgue point of $\chi_X$, then $\lim_{n\to\infty}\int_Xf_n(x)\,{\rm d}x=1$.\newline
(b) If $X\in\br$ is such that 
${\ell}(X)> 0$, then there is a point $a\in X$ such that
defining $g_n(x)=f_n(x-a)$ we have 
$\lim_{n\to\infty}\int_Xg_n(x)\,{\rm d}x=1$.
\end{theorem}

\begin{proof}
(a) By Lemma \ref{lemmaa} we may choose $\delta >0$ such that
${\ell}([0,r]\cap X)> r(1-\epsilon)$ and
${\ell}([-r,0)\cap X)> r(1-\epsilon)$
whenever $0<r\leq\delta$. By Lemma \ref{lemmab} we then have
$\int_{X\cap [0,\delta]}f_n(x)\,{\rm d}x\geq (1-\epsilon)\int_{[0,\delta]}f_n(x)\,{\rm d}x$,
and by an analogous argument we also get
$\int_{X\cap [-\delta,0)}f_n(x)\,{\rm d}x\geq (1-\epsilon)\int_{[-\delta,0)}f_n(x)\,{\rm d}x$.
Since $\lim_{n\to\infty}\int_{[-\delta,\delta]}f_n(x)\,{\rm d}x = 1$, there is $n_0\in\mathbb N$ such that
$\int_{[-\delta,\delta]}f_n(x)\,{\rm d}x>1-\epsilon$ whenever $n\geq n_0$.
Thus
\begin{eqnarray*}
\int_Xf_n(x)\,{\rm d}x&&\geq \int_{X\cap[-\delta,0)}f_n(x)\,{\rm d}x+ \int_{X\cap[0,\delta]}f_n(x)\,{\rm d}x\\
&& \geq (1-\epsilon)\left[  \int_{[-\delta,0)}f_n(x)\,{\rm d}x+ \int_{[0,\delta]}f_n(x) \,{\rm d}x \right]
\geq (1-\epsilon)^2
\end{eqnarray*}
for all $n\geq n_0$.\newline
(b) The Lebesgue points of $\chi_X$ form a set whose complement has measure zero
(see \cite[Theorem 7.7, p.\ 138]{Rudin}). Since $\ell(X)>0$, there is such a point $a\in X$. 
Using a translation, we may reduce the proof of this part to (a).
\end{proof}

\begin{remark}
In proving part (a) above the monotonicity conditions (i) and (ii) cannot be dispensed with.
For example, let
$$
X = \{0\}\cup\bigcup_{n=1}^\infty\left(\left[-\frac 1{2^n},-\frac 1{2^{n+1}}\left(1+\frac 1n\right)\right]
\cup\left[\frac 1{2^{n+1}}\left(1+\frac 1n\right),\frac 1{2^n}\right]   \right).
$$
It is easy to show that 0 is a Lebesgue point of $\chi_X$. On the other hand,
we can find (even continuous) functions $f_n:\mathbb R\to[0,\infty)$ such that
$\int_{\mathbb R}f_n(x)\,{\rm d}x=1$ and the support of $f_n$ is contained in the open interval
$\left(\frac 1{2^{n+1}},\frac 1{2^{n+1}} \left(1+\frac 1n\right)\right)$, so that (iv) holds but $\int_Xf_n(x)\,{\rm d}x=0$ 
for all $n\in\mathbb N$.
\end{remark}

We are now ready to derive a sufficient condition for a phase observable $E$
to satisfy $\|E(X)\|=1$ whenever $E(X)\ne O$.
Let $E$ be given with its phase matrix
$(c_{n,m})_{n,m\in\mathbb N}$.  For any unit vector $\psi\in\hi$, the phase probability measure $p^E_\psi$ 
is absolutely continuous with respect to the Lebesgue measure $\ell$.
Let $g^E_\psi$
denote the Radon-Nikod\'ym derivative,
so that 
$p_{\psi}^E(X)=(2\pi)^{-1}\int_Xg_\psi^E(\theta)\mathrm d\theta$.
This is a $2\pi$-periodic density function $\mathbb R\to[0,\infty]$.  
Consider  the following class of unit vectors
$$
\psi_r:=\sqrt{1-r^2}\sum_{n=0}^\infty e^{i\upsilon_n}r^n|n\rangle,\  \ r\in(-1,1),
$$
where $(\upsilon_n)_{n\in\mathbb N}\subset\mathbb R$.
The 
density function 
$g_{\psi_r}^E$ 
is continuous  and of the form
$$
g_{\psi_r}^E(\theta)=\left(1-r^2\right)\sum_{n,m=0}^\infty c_{n,m}e^{-i(\upsilon_n-\upsilon_m)}
r^{n+m}e^{i(n-m)\theta},
$$
where the series  converges absolutely. 

\begin{lemma}\label{lemmac}
With the above notations,
if $\lim_{n\to\infty}{c_{n,n+k}}e^{-i(\upsilon_n-\upsilon_{n+k})}=1$ for all $k\in\mathbb N$,
for some $(\upsilon_n)_{n\in\mathbb N}\subset\mathbb R$, then 
$$
\lim_{r\to1-}\frac{1}{2\pi}\int_{[-\delta,\delta]}
g_{\psi_r}^E(\theta)\mathrm d\theta=1
$$ 
for all $\delta>0$.
\end{lemma}

\begin{proof}
Suppose that 
\begin{equation}\label{xz}
\lim_{n\to\infty}{c_{n,n+k}}e^{-i(\upsilon_n-\upsilon_{n+k})}=1
\end{equation}
for all $k\in\mathbb Z^+$, where $(\upsilon_n)_{n\in\mathbb N}\subset\mathbb R$.
The Fourier-Stieltjes coefficients of the probability measure  $p^E_{\psi_r}$ are of the form
$$
c_k^r:=\frac{1}{2\pi}\int_0^{2\pi}e^{-ik\theta}g_{\psi_r}^E(\theta)\mathrm d\theta=
r^k\left(1-r^2\right)\sum_{n=0}^\infty c_{n+k,n}e^{-i(\upsilon_{n+k}-\upsilon_n)}r^{2n}
$$
and $c_{-k}^r=\overline{c_k^r}$ for all $k\in\mathbb N$ and $r\in(-1,1)$.
Next we show that $\lim_{r\to1-}c_k^r=1$ for all $k\in\mathbb Z$.

Fix $k\in\mathbb Z^+$ and $\epsilon>0$.
Since (\ref{xz}) holds, one may choose such an 
$n_\epsilon\in\mathbb Z^+$ that 
$\left|c_{n+k,n}e^{-i(\upsilon_{n+k}-\upsilon_n)}-1\right|<\epsilon/2$ for all 
$n\ge n_{\epsilon}$. Since $\sum_{n=0}^\infty r^{2n}=1\big/\left(1-r^2\right)$,
$r\in(-1,1)$, one gets
\begin{eqnarray*}
\left|c_k^r\big/r^k-1\right|&\le&\left(1-r^2\right)
\sum_{n=0}^{n_\epsilon-1}\left|c_{n+k,n}e^{-i(\upsilon_{n+k}-\upsilon_n)}-1\right|r^{2n}\\
&&+\;\left(1-r^2\right)\sum_{n=n_{\epsilon}}^\infty
\left|c_{n+k,n}e^{-i(\upsilon_{n+k}-\upsilon_n)}-1\right|r^{2n}\\
&\le&2\left(1-r^{2n_{\epsilon}}\right)+\epsilon/2.
\end{eqnarray*}
Choose $r_{\epsilon}\in[0,1)$ such that $2\left(1-r^{2n_{\epsilon}}\right)<\epsilon/2$
when $r\in[r_\epsilon,1)$ to get
$\left|c_k^r\big/r^k-1\right|<\epsilon$ for all $r\in[r_\epsilon,1)$. Thus, 
$c_k^r\sim r^{|k|}$ and $c_k^r\to1$ for all $k\in\mathbb Z$ when $r\to1-$.

The condition $\lim_{r\to1-}c_k^r=1$, $k\in\mathbb Z$, implies that 
$$
\lim_{r\to1-}\frac{1}{2\pi}\int_{-\delta}^{\delta}
g_{\psi_r}^E(\theta)\mathrm d\theta=1
$$ 
for all $\delta>0$.
\end{proof}

Lemma \ref{lemmac} applies, in particular, to the canonical phase.
Moreover,
in that case the density function $g_{\psi_r}^{E_{\rm can}}$ is simply
$$
g_{\psi_r}^{E_{\rm can}}(\theta)=\left(1-r^2\right)\sum_{n,m=0}^\infty r^{n+m}e^{i(n-m)\theta}
=\frac{1-r^2}{1-2r\cos\theta+r^2}.
$$
Defining
$f_{n}(x):=g_{\psi_{1-(n+1)^{-1}}}^{E_{\rm can}}(x)$, $|x|\le\pi$, and
$f_{n}(x)=0$, $|x|>\pi$, one gets a sequence $(f_n)_{n\in\mathbb N}$
which fulfills the conditions of Theorem \ref{lausea}.
%
%
%
Consider next the phase observable $E_{|0\rangle}$. For  coherent states
$|r\rangle = e^{-r^2/2}\sum_{n\geq 0} \frac{r^n}{\sqrt{n}}|n\rangle$, $r\geq 0$, we obtain
\cite{LP00}
$$
g^{E_{|0\rangle}}_{|r\rangle}(\theta)= \int_{r^2}^\infty e^{-v}\,{\rm d}v
+ e^{-r^2\sin^2\theta}\,2r\cos\theta\int_{-r\cos\theta}^\infty e^{-u^2}\,{\rm d}u.
$$
Functions 
$f_{n}(x):=g^{E_{|0\rangle}}_{|r=n\rangle}(x)$, $|x|\le\pi$, $f_n(x)=0$, $|x|>\pi$, 
also fulfill the conditions of Theorem \ref{lausea}. Hence we have the following results:
\begin{proposition}
If $X\in\mathcal B([0,2\pi))$ has nonzero Lebesgue measure, then
 $\parallel E_{\rm can}(X)\parallel =1$
and $\parallel E_{|0\rangle}(X)\parallel =1$.
\end{proposition}

Hence both the canonical phase $E_{\rm can}$ and the ground state phase $E_{|0\rangle}$
fulfill this necessary condition for the number-phase complementarity.
The question remains, however,  whether these observables actually are complementary 
to the number.


\section{On number-phase uncertainty relations}

The number-phase uncertainty relations are often presented
as a kind of quantitative expression for the  complementarity of this
pair of observables. Although we do not support this viewpoint,
we find it useful to briefly elaborate on the number-phase uncertainty
product, especially for high amplitude coherent states.

A phase observable $E$ is a periodic quantity. Therefore, the variance 
${\rm Var}\,(E,\psi)$ of the phase distribution $p^E_\psi$ in a vector state 
$\psi\in\hi$, $\parallel\psi\parallel=1$, though well defined, is not a good measure of phase
uncertainty.  For periodic distributions the appropriate notion is that of 
minimun variance, introduced by L\'evy \cite{Levy}. Using the density $g^E_\psi$
of $p^E_\psi$, 
the {\em minimum variance} of the phase distribution $p^E_\psi$ is then defined as
$$
{\rm VAR}\,(E,\psi) := \inf\left\{\frac 1{2\pi}\int_{\beta-\pi}^{\beta+\pi}(\theta-\alpha)^2\,g^E_\psi(\theta)\,d\theta
\,\Big|\, \alpha,\beta\in\mathbb R\right\},
$$
and one finds that $0\leq{\rm VAR}\,(E,\psi)\leq \pi^2/3$.
The minimum variance of the canonical phase $E_{\rm can}$ in a coherent state
$\ket z$ has, for large $|z|$, the following asymptotic form
(for details, see \cite[VII A,C]{LP00}):
$$
{\rm VAR}\,(E_{\rm can},\ket z) \simeq \frac 14\frac 1{|z|^2}.
$$
On the other hand, the variance of the number observable $N$ in a coherent state $\ket z$ is 
${\rm Var}\,(N,\ket z) = |z|^2$. so that, for large $|z|$, one has
$$
{\rm Var}\,(N,\ket z)\,{\rm VAR}\,(E_{\rm can},\ket z) \simeq \frac 14.
$$

The logical independence of complementarity and uncertainty relations in
general has been clearly established long ago \cite{Lahti}. The concept
of complementarity is linked with the impossibility of joint
measurements of two observables. By contrast, as is evident from the
above formalisations, the uncertainty relation,
as well as the notions of probabilistic and value
complementarity, refer to features of the probability distributions 
of separate, independent measurements of the two observables in question.

\section{Conclusion}

The set of covariant phase observables is a convex set  in a natural way.
Let $E$ be any phase observable.
Then  
$E_\epsilon(X) = \epsilon E_{\rm triv}(X) + (1-\epsilon)E(X)$, $X\in\mathcal B\left([0,2\pi)\right)$,
defines a phase observable which is not complementary with number.
Indeed, $E_\epsilon(X)\geq \epsilon P_k$ for all $k\geq 0$ so that
$E_\epsilon(X)\land P_k\geq \epsilon P_k$ for all $k\geq 0$.
This shows that every phase observable $E$  is ``arbitrarily close" to a phase observable
$E_\epsilon$ which is not complementary with the number. 

This observation, which generalises to every canonical pair, implies that the complementarity 
of such pairs, given that it holds, is not strictly testable. This is also true for probabilistic 
complementarity in the sense that finite statistics can never confirm strictly whether a given event 
has probability equal to one. Nevertheless, complementarity indicates a relation between two 
observables which is robust under small imprecisions: if a pair of 
observables is complementary, then a ``nearby" noncomplementary pair will only allow ``small" 
positive joint lower bounds between their positive operators.

The canonical phase, as well as any unitarily equivalent one, is an extremal element
of the convex set of phase observables. Indeed,
by Theorem~\ref{uni}, for all such phase observables $|c_{n,m}|=1$,  $n,m\geq 0$. 
Let $(c_{n,m})$ be the phase matrix of a phase observable $E$, and assume that
$E$ is a convex combination of phase observables $E_1$ and $E_2$, that is,
$c_{n,m}= \lambda c^{(1)}_{n,m} +(1-\lambda)c^{(2)}_{n,m}$,
with some $0<\lambda<1$. 
If $|c_{n,m}|=1$, 
it follows that $| c^{(1)}_{n,m}|$=$| c^{(2)}_{n,m}|$=1, and that the phases of
$c^{(1)}_{n,m}$ and $c^{(2)}_{n,m}$ are the same. Therefore, $c_{n,m}= c^{(1)}_{n,m} =c^{(2)}_{n,m}$
for all $n,m\geq 0$, that is, $E=E_1=E_2$.
We conclude with the conjecture that further analysis of the convex
structure of the set of phase observables may help to decide on the open
question of the existence of a phase that is complementary to number.



\end{document}